\documentclass[aps,prl,
              reprint,
              amsmath,amssymb,amsfonts,groupedaddress,
              longbibliography,
              superscriptaddress,floatfix]{revtex4-1}

\usepackage{pdfpages}

\makeatletter
\AtBeginDocument{\let\LS@rot\@undefined}
\makeatother

\usepackage{graphicx}
\usepackage{dcolumn}
\usepackage{bm}
\usepackage{units}
\usepackage{upgreek}
\newcommand{\nfrac}[2]{{#1}/{#2}}

\usepackage{color}

\usepackage{hyperref}
  \hypersetup{
  colorlinks=true,
  linkcolor=blue,
  citecolor=green,
  filecolor=black,
  menucolor=black,
  urlcolor=blue
  }
 
\usepackage{breakurl}

\newcommand{\sdist}{\kern 0.20em}
\renewcommand{\eqref}[1]{Eq.\sdist(\ref{#1})}
\newcommand{\figref}[1]{Fig.\sdist\ref{#1}}

\DeclareMathOperator{\Kbessel}{K}

\newcommand{\average}[1]{\left\langle#1\right\rangle}

\allowdisplaybreaks

\newcommand{\eps}{\varepsilon}
\newcommand{\Wglauber}{\mathcal{W}}
\usepackage{upgreek}

\begin{document}

\title{Efficient High-Energy Photon Production in the Supercritical QED Regime}

\author{Matteo Tamburini}\email{matteo.tamburini@mpi-hd.mpg.de}
\affiliation{Max-Planck-Institut f\"ur Kernphysik, Saupfercheckweg 1, D-69117 Heidelberg, Germany}
\author{Sebastian Meuren}\email{smeuren@stanford.edu}
\affiliation{Stanford PULSE Institute, SLAC National Accelerator Laboratory, Menlo Park, CA 94025}
\affiliation{Department of Astrophysical Sciences, Princeton University, Princeton, NJ 08544}

\date{\today}

\begin{abstract}
When dense high-energy lepton bunches collide, the beam particles can experience rest-frame electromagnetic fields which greatly exceed the QED critical one. Here it is demonstrated that beamstrahlung efficiently converts lepton energy to high-energy photons in this so-called supercritical QED regime, as the single-photon emission spectrum exhibits a pronounced peak close to the initial lepton energy.  
It is also shown that the observation of this high-energy peak in the photon spectrum requires one to mitigate multiple photon emissions during the interaction. Otherwise, the photon recoil induces strong correlations between subsequent emissions which soften the photon spectrum and suppress the peak. The high-energy peak in the photon spectrum constitutes a unique observable of photon emission in the supercritical QED regime, and provides decisive advantages for the realization of an efficient multi-TeV laserless gamma-gamma collider based on electron-electron collisions. 
\end{abstract}

\maketitle

A future multi-TeV lepton collider has to be linear in order to mitigate energy losses via synchrotron radiation~\cite{shiltsevRMP21}. As two colliding bunches cross only once in a linear collider, extremely high particle densities are necessary at the interaction point in order to achieve the luminosities required to search for physics beyond the standard model~\cite{tanabashiPRD18, badelek2004, Lebrun:1475225}. As a result, beamstrahlung energy losses become a decisive limiting factor, especially in the multi-10-TeV regime~\cite{nobleNIMA87, telnovNIMA90, chenPRD92, yokoya_beam-beam_1992, esberg_strong_2014}. 

Beamstrahlung is primarily characterized by the quantum parameter $\chi = \Upsilon = F^*/F_{\text{cr}}$ \cite{yokoya_beam-beam_1992}, where $F^*$ denotes the electric field in the rest frame of a beam particle and $F_{\text{cr}}=m^2 c^3/|e| \hbar \approx 1.3\times 10^{18}\,\text{V/m}$ is the QED critical (Schwinger) field $(F_{\text{cr}}/c \approx 4.4 \times 10^{9}\,\text{T})$. For $\chi\ll 1$ the radiative energy loss of an unbound ultrarelativistic charge is well approximated by the prediction of classical electrodynamics, whereas for $\chi \gtrsim 1$ quantization effects in the radiation field become essential~\cite{ritusJSLR85, Baier-book, dipiazzaRMP12}. While the regime $\chi\lesssim 1$ is relatively well explored theoretically~\cite{ritusJSLR85, dipiazzaRMP12} (see~\cite{yan_high-order_2017, colePRX18, poderPRX18, wistisenNC18} for recent experiments), the supercritical regime $\chi \gg 1$ is still poorly understood. When $\alpha \chi^{2/3} \gtrsim 1$ ($\chi \gtrsim 10^3$), where $\alpha = e^2/(4\pi\epsilon_0\hbar c) \approx 1/137$ is the fine-structure constant, radiative corrections become significant~\cite{ritusJSLR85}, and even a complete breakdown of perturbation theory has been conjectured~\cite{ritus70, narozhnyPRD80} (see also~\cite{fedotovJPCS17, podszusPRD19, ildertonPRD19, dipiazzaPRD20, mironov_resummation_2020} for recent theoretical studies and~\cite{yakimenkoPRL19, blackburnNJP19, baumannSR19, baumannPPCF19, dipiazzaPRL20} for proposals to probe this regime experimentally). 

State-of-the-art linear lepton collider designs such as CLIC and ILC~\cite{aicheler_compact_2019,bambade_international_2019} employ long and flat bunches in order to minimize beamstrahlung. Recently, it was suggested in Ref.~\cite{yakimenkoPRL19} that beamstrahlung could also be mitigated by colliding short and round bunches and operating in the supercritical QED regime ($\chi\gg 1$). A different approach, which completely circumvents the beamstrahlung problem, are gamma-gamma colliders~\cite{ginzburg_production_1981, telnovNIMA90, gronbergRAST14, takahashi_future_2019}. 

The state-of-the-art concept to generate high-energy photons for a gamma-gamma collider is based on Compton backscattering of two intense laser pulses with two counterpropagating lepton bunches, properly coordinated in space and time~\cite{telnovNIMA90, telnovNIMA95, gronbergRAST14, takahashi_future_2019}. However, Compton backscattering becomes increasingly more challenging to realize with increasing center-of-mass energy~\cite{telnovNIMA90, takahashi_future_2019}. Remarkably, beamstrahlung itself could be used to produce high-energy gamma photons~\cite{blankenbeclerPRL88, delgaudioPRAB19}. 

Here we point out that qualitatively new features appear in the photon emission spectrum in the supercritical QED regime. In fact, we demonstrate that (i)~for $\chi \gtrsim 16$ the probability for an electron to emit a single photon carrying almost all its initial energy strongly increases; (ii)~the single-emission photon spectrum can be observed in asymmetric electron-electron beam collisions and provides direct quantitative \emph{in situ} information on the average $\chi$ of the beam achieved during the collision; (iii) when multiple photon emissions become dominant, the high-energy peak in the total and, remarkably, even in the single-photon emission spectrum vanishes. 

On one hand, these findings are of intrinsic interest for both fundamental science and practical applications. In fact, they provide a unique observable which, on a shot-to-shot basis, directly informs on the actually achieved average $\chi$ of the beam, therefore overcoming experimental uncertainties such as jitter in the collision parameters. On the other hand, they reveal an optimal regime for efficient high-energy photon production. This optimal regime is attained by properly shaping the colliding bunches such that most of the electrons of the beam reach $\chi\gg1$ and emit once, while further emissions by the same beam electrons remain negligible. This allows one to greatly increase the yield of photons that carry nearly all the energy of the emitting electron while simultaneously suppressing the low-energy photon background. This is a decisive advantage for a gamma-gamma collider as, during gamma-gamma collisions, low-energy photons can convert high-energy photons into $e^-e^+$ pairs via the linear Breit-Wheeler process,  therefore substantially reducing the luminosity of high-energy photons.

\begin{figure}[tb]
\centering
\includegraphics[width=\linewidth]{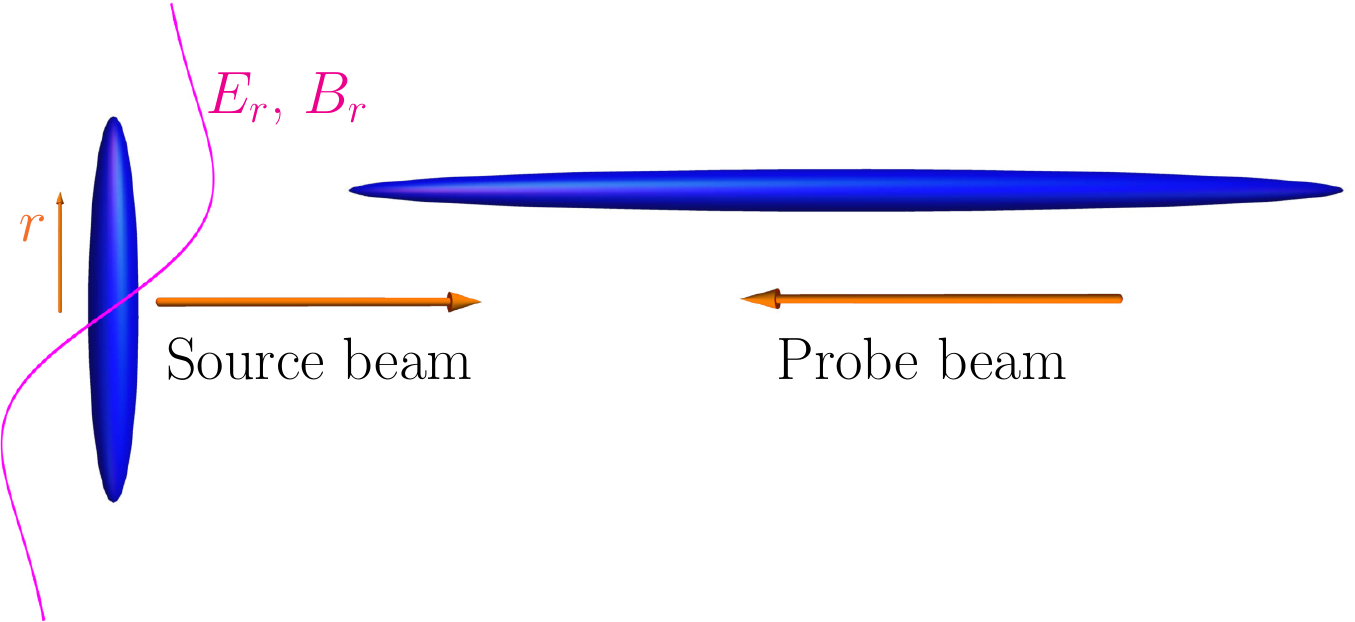}
\caption{Schematic setup. A pancake-shape dense source beam collides with an elongated cigar-shape low-density probe beam. The probe beam collides with a transverse impact parameter $r$, such that the source beam electric $E_r$ and magnetic $B_r$ fields approach their maximum.}
\label{fig:1}
\end{figure}

In the following we consider an asymmetric electron-electron collider setup (see \figref{fig:1}). A short dense pancake-shape electron ``source'' beam collides head on with an elongated cigar-shape high-energy ``probe'' beam. For clarity, source and probe-beam parameters are denoted with the subscript $s$ and $p$, respectively. As the electromagnetic field experienced by the probe beam particles changes significantly as a function of the impact parameter, the considered asymmetric setup avoids a trivial average over different values of $\chi_p$, which is always present in symmetric collisions. In comparison with symmetric collisions the ``source'' beam provides the strong field, i.e., its longitudinal bunch length ($\ell_s$), transverse rms size ($\sigma_s$), and number of electrons ($N_s$) are relevant for calculating the field. On the other hand, the high-energy ``probe'' beam provides the large gamma factor $\gamma_p$ to boost the experienced rest-frame field. Note that the energy of the source beam is not relevant for attaining large $\chi_p = \gamma_p \sqrt{(\bm{E}_s + \bm{v}_p \times \bm{B}_s)^2-(\bm{E}_s\cdot\bm{v}_p/c)^2}/F_{\text{cr}}$, where $\bm{E}_s$ ($\bm{B}_s$) is the electric (magnetic) field of the source beam and $\bm{v}_p$ ($\gamma_p$) is the velocity (relativistic factor) of the probe beam in the laboratory frame.  In fact, the quantum parameter $\chi_p$ can also be expressed directly in terms of the beam parameters \cite{yokoya_beam-beam_1992}
\begin{gather}
	\label{eqn:chi_scaling}
	\chi_p \sim \gamma_p \alpha N_s \frac{\lambdabar_c^2}{\sigma_s\ell_s}, \text{ with } \lambdabar_c = \frac{\hbar}{mc} \approx \unit[3.9 \times 10^{-13}]{m}.
\end{gather}
For the beam parameters considered here, the source beam remains almost unaffected by the interaction with the probe beam (see below and Ref.~\cite{yokoya_beam-beam_1992}). Furthermore, the interaction is collisionless, i.e., the probe beam interacts only with the collective electromagnetic field of the source beam. As the probe beam is ultrarelativistic, the quasiclassical approximation holds, i.e., particle trajectories are calculated using classical electrodynamics, whereas the emission itself is calculated quantum mechanically~\cite{ritusJSLR85,Baier-book,dipiazzaRMP12}. The formation length for photon emission scales as (assuming $\chi \gg 1$)~\cite{baier_quantum_1989}
\begin{gather}
l_f \sim \frac{\gamma \lambdabar_c}{u^{1/3}\chi^{2/3}}, \quad u = \frac{\omega}{1 - \omega}, \quad \omega = \frac{\varepsilon_\gamma}{\varepsilon}.
\end{gather}
For photon energies $\varepsilon_\gamma$ comparable to the initial energy $\varepsilon = \gamma mc^2$ of the emitting electron ($u \gtrsim 0.1$) and for the beam parameters considered here ($\gamma \sim 10^5$, $\chi \sim 100$, $\ell_s \sim \unit[100]{nm}$), the formation length is much shorter than the scale on which the electromagnetic field of the source beam changes ($l_f \ll \ell_s$). Furthermore, the electron energy is negligibly altered during the emission process as $|e \bm{E}_s| l_f \sim mc^2 (\chi/u)^{1/3} \ll \gamma mc^2$. Thus, the source-beam field is locally constant during the photon emission process~\cite{dipiazzaPRA18, ildertonPRA19, dipiazzaPRA19}, which allows one to employ the differential radiation probability in a constant homogeneous field, where the local value of $\chi$ and $\gamma$ is used~\cite{ritusJSLR85, yokoya_beam-beam_1992, Baier-book}:
\begin{multline} \label{dW}
\frac{d^2W}{dt\,d\omega}  = \frac{\alpha}{\sqrt{3} \pi \tau_c \gamma} \bigg\{\bigg[ 2 + \frac{\omega^2}{(1-\omega)}\bigg] \Kbessel_{2/3}\bigg[\frac{2 \omega}{3 \chi (1-\omega)}\bigg] \\
 - \int^{\infty}_{2 \omega / [3 \chi (1-\omega)]}{dy\,\Kbessel_{1/3}(y)}\bigg\}.
\end{multline}
Here $\tau_c=\hbar/(mc^2)$ is the Compton time, and $\Kbessel_{\nu}(x)$ is the modified Bessel function of the second kind~\cite{olver_nist_handbook_2010}. 
Correspondingly, the local emission rate is $\nfrac{dW}{dt} = \int_0^1{d\omega\, \nfrac{d^2W}{dt d\omega}}$, while the normalized emitted power is $\nfrac{dI}{dt} = \int_0^1 d\omega\, \nfrac{d^2I}{dt\,d\omega}$, where $\nfrac{d^2I}{dtd\omega} = \omega \nfrac{d^2W}{dtd\omega}$.

\begin{figure}[tb]
\centering
\includegraphics[width=\linewidth]{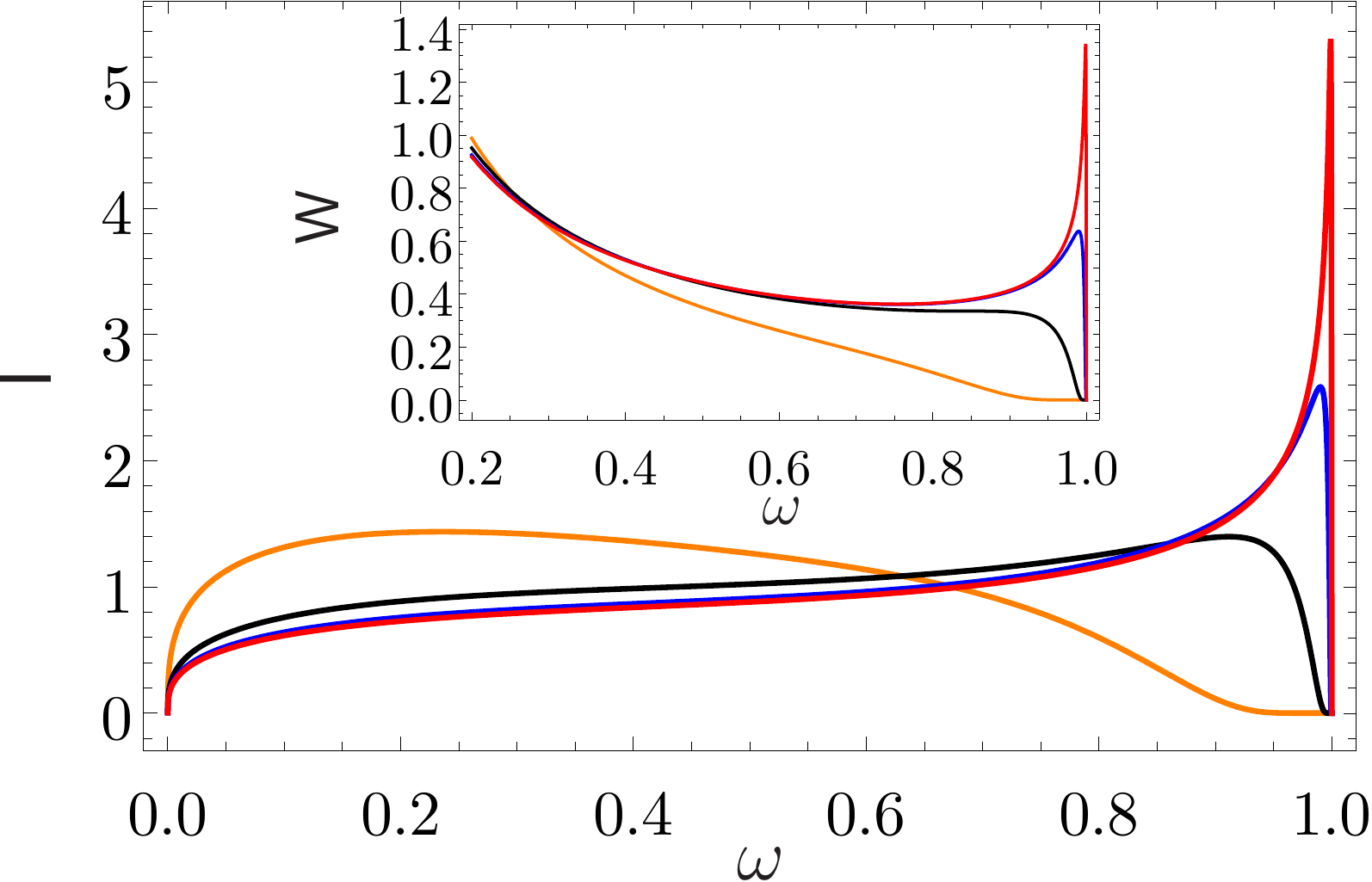} 
\caption{Normalized photon emission spectrum $\mathsf{I} = \nfrac{d^2I}{dt\,d\omega}(\nfrac{dI}{dt})^{-1}$ and normalized probability $\mathsf{W}= \nfrac{d^2W}{dt\,d\omega}(\nfrac{dW}{dt})^{-1}$ (inset) for $\chi=1.6$ (orange lines), $\chi=16$ (black lines), $\chi=160$ (blue lines), and $\chi=1600$ (red lines).}
\label{fig:2}
\end{figure}

A detailed analysis of \eqref{dW} reveals a distinctive feature of the photon emission spectrum which occurs exclusively in the supercritical regime. Whereas $d^2W/dtd\omega$ is a monotonically decreasing function of $\omega$ for $\chi \lesssim 16$, it develops a local minimum and maximum for $\chi \gtrsim 16$, which results in a peak close to $\omega=1$ (see inset of \figref{fig:2} and Refs.~\cite{esberg_does_2009, bulanovPRA13}). For $\chi > 16$ the minimum and maximum are approximately located at, respectively,
\begin{gather} \label{minmax}
\begin{gathered}
\omega_{\text{min}}  \approx 0.754 + \frac{(15.7 + 0.146 \chi)}{\chi^2},\\
\omega_{\text{max}}  \approx 1 - \frac{(174 + 20 \chi)}{15 \chi^2}.
\end{gathered}
\end{gather}
The height of the peak $H$ is given by
\begin{gather} \label{H}
H = \frac{(d^2W/dtd\omega)(\omega_{\text{max}})}{(d^2W/dtd\omega)(\omega_{\text{min}})}   \approx \frac{1.315 + 0.315 \chi}{\chi ^{2/3}}
\end{gather}
and provides a unique observable of the average $\chi$ that is actually achieved during the interaction. Note that the peak at $\omega\approx1$ originates from the factor $(1-\omega)^{-1}$, and guarantees that \eqref{dW} conserves energy even in the deep quantum regime $\chi\gg 1$. In fact, photon emission with $\varepsilon_\gamma > \varepsilon$ is impossible within the approximation that the energy transferred by the field during the emission itself is negligible.

Figure\sdist\ref{fig:2} displays the normalized emission spectrum  $\mathsf{I} = \nfrac{d^2I}{dt\,d\omega}(\nfrac{dI}{dt})^{-1}$ and the normalized photon emission probability $\mathsf{W}= \nfrac{d^2W}{dt\,d\omega}(\nfrac{dW}{dt})^{-1}$ in four different regimes: (i)~the critical regime ($\chi\sim 1$, orange line), (ii)~transition between the critical and the supercritical regime ($\chi\sim 10$, black line), (iii)~the supercritical regime ($\chi\sim 100$, blue line), and (iv)~the fully nonperturbative regime ($\alpha \chi^{2/3}\sim 1$, red line). Whereas electrons still emit in a broad energy range for $\chi\sim1$, a sharp peak close to the initial electron energy ($\omega\approx 1$) appears in the supercritical regime ($\chi>16$). The probability of producing a photon with energy beyond $\omega_{\text{min}}$ already exceeds 9\% for $\chi>60$ and basically saturates to approximately 11\% for $\chi\gtrsim 800$. However, the height of the peak monotonically increases with increasing $\chi$; i.e., the photon spectrum at $\omega\approx1$ becomes more and more monochromatic [see \eqref{H} and \figref{fig:2}].

In order to quantitatively investigate the photon emission spectrum, 3D Monte Carlo simulations of beam-beam collision were performed using the state-of-the-art methodology~\cite{rimbault_guinea-pig_2007, gonoskovPRE15, tamburiniSR17}. In the simulations an exact 3D analytical solution of Maxwell's equations was used for the electromagnetic fields of the source beam (see Supplemental Material\footnote{See Supplemental Material for (i) exact three dimensional (3D) analytical solution of Maxwell's equations for a finite size charged beam; (ii) details of the employed Monte Carlo simulation technique; (iii) details of the derivation of the multiphoton emission distribution.}). The source beam has $\unit[10]{GeV}$ energy, $\unit[0.96]{nC}$ charge, $\ell_s=\unit[100]{nm}$ bunch length (in the laboratory frame), and $\sigma_s=\unit[300]{nm}$ transverse size. In the laboratory, a maximum electric (magnetic) field of $E_{\text{max}}\approx \unitfrac[2.6 \times 10^{14}]{V}{m}$ ($B_{\text{max}}\approx \unit[8.6 \times 10^{5}]{T}$) is achieved at an impact parameter of $r \approx \unit[500]{nm}$. The source beam parameters considered here are comparable to those achievable at the FACET-II facility at SLAC~\cite{yakimenko_facet-ii_2019}. 

The probe beam has $\unit[100]{GeV}$ energy with $\unit[100]{MeV}$ rms energy spread, $\unit[16]{pC}$ charge, $\ell_p=\unit[300]{\upmu{}m}$ bunch length, and $\sigma_p=\unit[50]{nm}$ transverse size. As a result, the maximum electric (magnetic) field of the probe beam $E_{\text{max}}\approx \unitfrac[8.7 \times 10^{9}]{V}{m}$ ($B_{\text{max}}\approx \unit[29]{T}$) at $r \approx \unit[80]{nm}$ is much weaker and its density is much lower than the corresponding values of the source beam. Thus, the source beam is basically unaffected by the interaction as both collisions and energy losses associated with the probe beam fields are negligible. Due to the finite transverse size of the probe beam $\chi_p$ ranges approximately from 75 to 77. For the above parameters the average emission probability per electron is approximately $0.12$. Note that nanometer-scale beam stabilization and thus smaller than 1\%-level fluctuations in $\chi$ are anticipated for state-of-the-art final focusing systems \cite{redaelli_stabilization_2003, yan_measurement_2014, marin_final_2018}. In addition, each electron emits on average less than once and photon emission occurs in a cone with $1/\gamma_p \approx \unit[5]{\upmu{}rad}$ opening angle around the propagation direction of the emitting electron~\cite{Baier-book, blackburn_radiation_2020}. Hence, the angular distribution of the emitted photons is to excellent accuracy the same as the electron beam angular distribution.

\begin{figure}[tb]
\centering
\includegraphics[width=\linewidth]{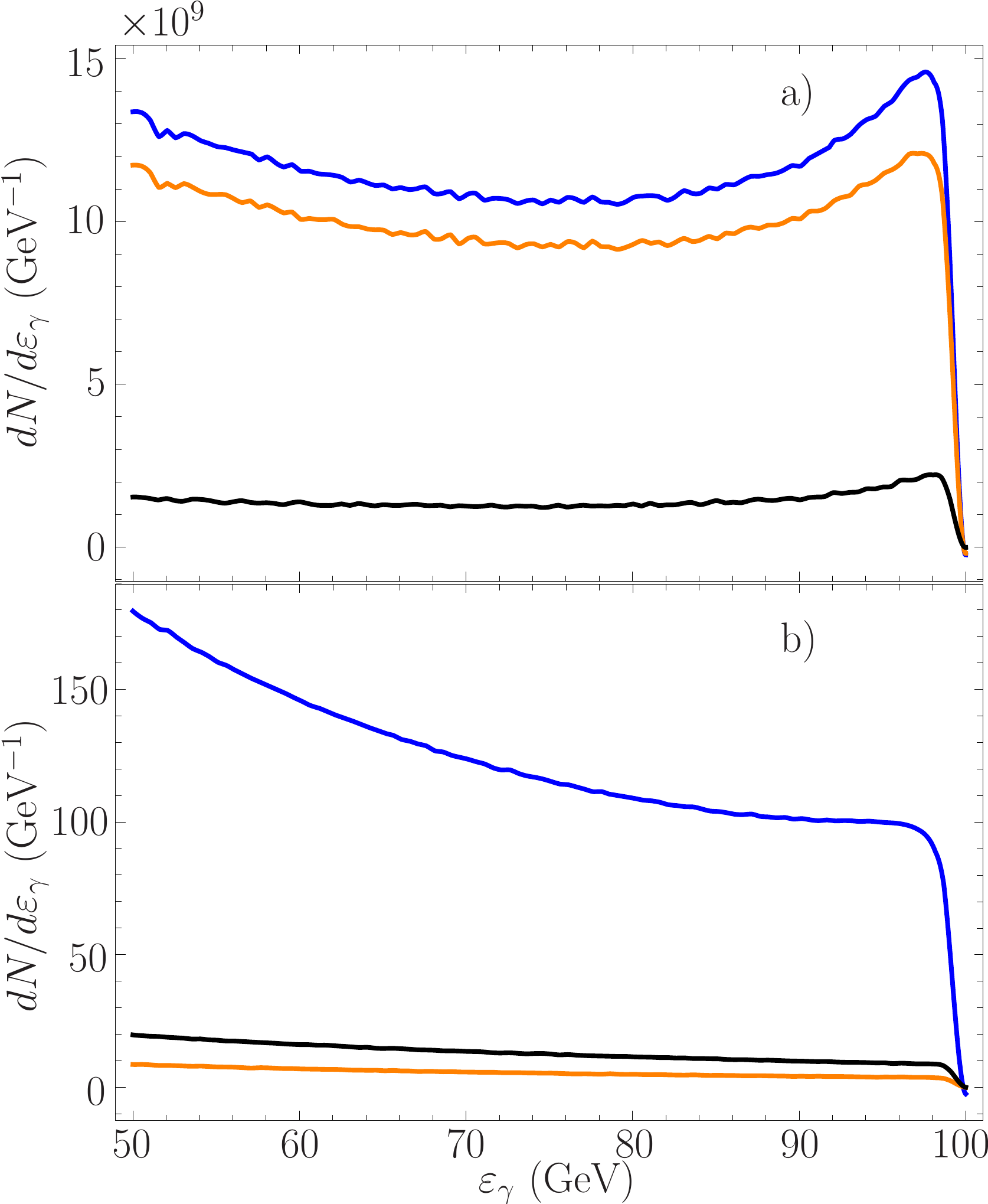}
\caption{Photon emission probability. The solid orange, black and blue lines report photons that originate from electrons that emitted only once, only twice, and all final photons, respectively. \textbf{(a)} Short interaction time ($\ell_s = \unit[100]{nm}$), single emissions dominant \textbf{(b)} long interaction time ($\ell_s = \unit[2500]{nm}$, multiple emissions dominant. See the main text for further details.}
\label{fig:3}
\end{figure}

\begin{figure}[tb]
\centering
\includegraphics[width=\linewidth]{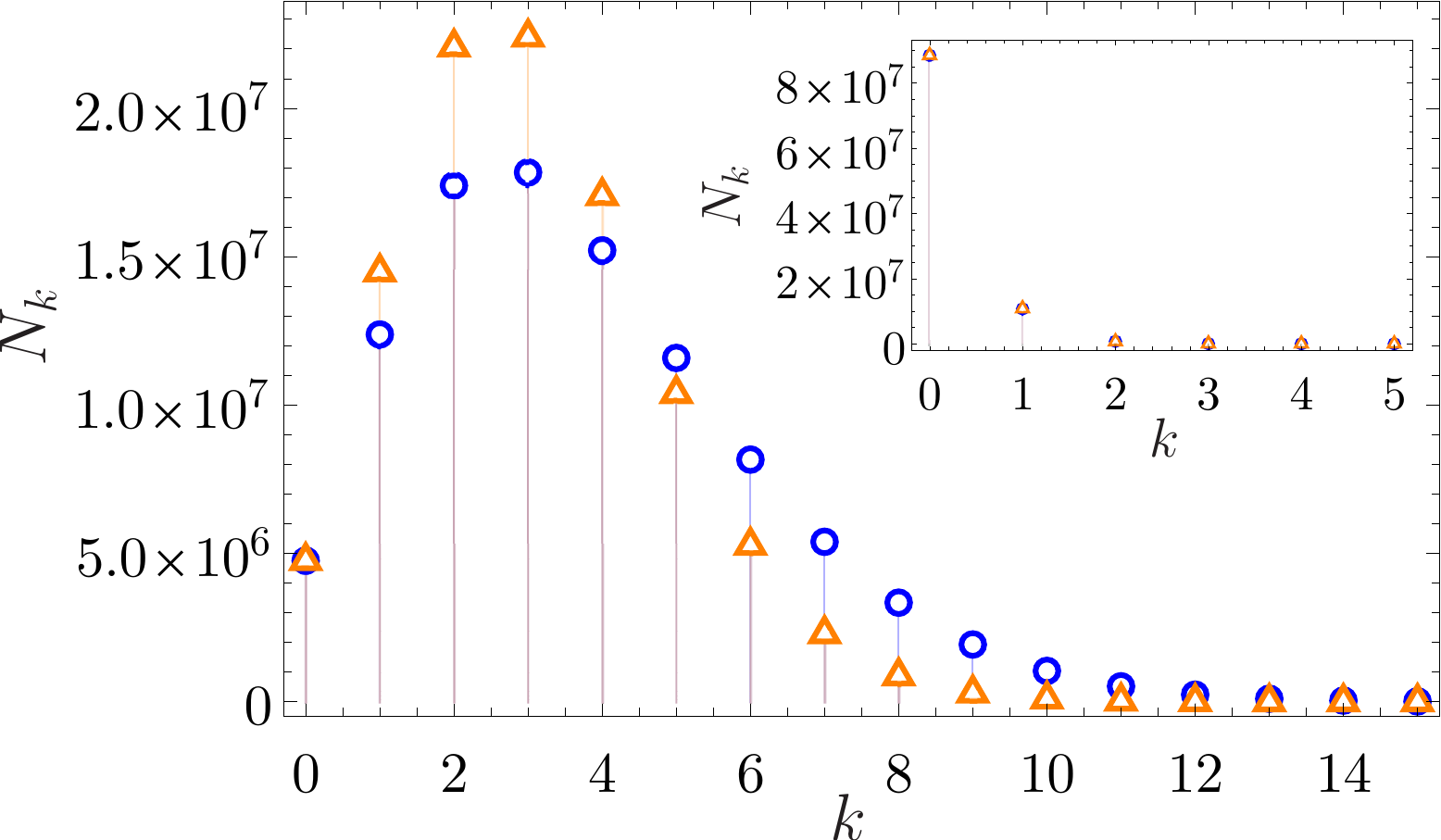}
\caption{Number of probe beam electrons $N_k$ that emitted $k$ photons in the interaction with the source beam. Blue circles report the simulation results, orange triangles the Poisson prediction. Main plot: $\ell_s = \unit[2500]{nm}$. Inset: $\ell_s = \unit[100]{nm}$.}
\label{fig:4}
\end{figure}

Figure\sdist\ref{fig:3}(a) reports the photon energy distribution for electrons which emitted \emph{only one} (\emph{two}) photon(s) during the interaction [orange (black) line] and the total photon distribution (blue line). Accordingly, the photon spectrum is dominated by single emissions while secondary and higher-order processes are suppressed (see also the inset of \figref{fig:4}). The peak height $H$ obtained from simulations (total spectrum, blue line) corresponds to $\chi_p \approx 69$ if \eqref{H} is employed. This value is within the 10\% error margin which we expect due to the presence of multiple emissions. In fact, $H$ provides a lower bound to $\chi_p$, which converges to the actual value in the single-emission limit. 

Next, we consider the same parameters as above but increase (decrease) the source beam length (transverse size) by a factor of $25$, i.e., employ $\ell_s=\unit[2500]{nm}$ and $\sigma_s=\unit[12]{nm}$. This scaling leaves $\chi_p$ invariant [see \eqref{eqn:chi_scaling}], which is now reached at an impact parameter $r\approx\unit[20]{nm}$. To keep the variation of $\chi_p$ comparable to the first simulation, we also reduce the transverse size of the probe beam to $\sigma_p=\unit[2]{nm}$. As a result, the average number of photon emissions per electron increases to approximately $3.0$ (see \figref{fig:3}b and \figref{fig:4}). In contrast to the previous simulation, the spectrum no longer exhibits a peak. 

In order to explain this transition, we assume that a probe particle experiences the supercritical quantum regime ($\chi_p\gg 1$). As shown in \figref{fig:2}, it is likely that this particle emits a hard photon with $\omega \approx 1$. Due to the large recoil, the probe particle has a much lower energy $\varepsilon' = \eps(1-\omega)$ after the emission. In the regime $\chi_p\gg 1$, the scaling of the radiation probability $W_p \sim [\ell_s / (\lambdabar_c \gamma_p)]^{1/3}$ (see, e.g., Refs.~\cite{yokoya_beam-beam_1992, ritusJSLR85, yakimenkoPRL19}) implies that a particle with lower energy has an increased radiation probability. Therefore, the emission of a hard photon ($\omega \approx 1$) increases the probability to emit a second photon, which is on average much softer. On the contrary, the emission of a soft photon ($\omega \ll 1$) is less likely followed by a second emission. As a consequence, we expect the peak to vanish in the total photon spectrum when multiphoton emissions are dominant. Remarkably, the peak disappears also in the one-photon emission spectrum (see the orange line in \figref{fig:3}b), which is naively not expected based on perturbation theory (see below).

In the following the photon emission distribution is calculated analytically.  Assuming that the locally-constant-field approximation holds, the single-photon emission probability given in \eqref{dW} is always applicable for sufficiently small time intervals $dt$. Thus, the probability $S(t,t';\eps)$ that an electron with energy $\eps$ does \emph{not} emit a photon during the time interval $[t,t']$ is given by (see Supplemental Material)
\begin{gather}
\label{eqn:decayfactor}
S(t,t';\eps) = \exp \bigg[ -\int_{t'}^{t} d\tau \, \frac{dW}{d\tau}(\eps,\tau) \bigg].
\end{gather}
Correspondingly, the electron ``decays'' exponentially, with a radiative lifetime given by the total emission probability per unit time $\nfrac{dW(\eps,t)}{dt} = \int_0^{\eps} d\eps' \, \nfrac{d^2W(\eps',\eps,t)}{dt\,d\eps'}$. As $\eps' = \eps(1-\omega)$, the quantities ${d^2W}/{dtd\eps'}$ and ${d^2W}/{dtd\omega}$ are trivially related [see \eqref{dW}]. Note that $S(t,t';\eps)$ can also be derived from the radiatively corrected wavefunction of the electron (see \cite{meurenPRL11, meuren_diss_2015, dinuPRD12, podszusXXX21} for further details).

The probability $(\nfrac{dP_1}{d\eps'})(\eps',t)$ that an electron has an energy within $[\eps',\eps'+d\eps']$ at time $t$ after radiating exactly one photon is
\begin{multline}
\label{eqn:dP_1depsprime}
\frac{dP_1}{d\eps'}(\eps',t) =  \int_{-\infty}^{t} d\tau \, S(t,\tau;\eps')
\\ 
\times  \frac{d^2W}{d\tau\,d\eps'}(\eps',\eps_i,\tau) S(\tau,-\infty;\eps_i).
\end{multline}
Here and in the following, $\eps_i$ denotes the initial electron energy, we implicitly assume that the work performed by the external field is negligible compared to the electron energy, and $\chi(t)$ is obtained from the electron trajectory. Equation~(\sdist\ref{eqn:dP_1depsprime}) explains the suppression of the peak even in the one-photon emission spectrum when multiple emissions become dominant (see Fig.~\ref{fig:3}). In fact, in the regime $\chi\gg 1$ substantial recoil is likely, which implies $\eps' \ll \eps_i$. Correspondingly, the decay exponent after the emission $S(t,\tau;\eps')$ is substantially smaller than it would be with negligible recoil $S(t,\tau;\eps'\approx\eps_i)$, i.e., the electron ``radiative lifetime'' substantially decreases after the emission. Therefore, the high-energy part of the spectrum $\omega = 1 - \eps'/\eps_i \approx 1$ is suppressed for sufficiently long interaction time. Consequently, even the one-photon emission spectrum differs qualitatively from Eq.~(\ref{dW}) [compare Figs.~\figref{fig:3}(a) and \figref{fig:3}(b)]. Note that for short interaction times $S(t,\tau;\eps') \approx 1$, independently of the magnitude of the recoil [see \eqref{eqn:decayfactor}], and the spectrum coincides with Eq.~(\ref{dW}).

Equation~(\sdist\ref{eqn:dP_1depsprime}) can be easily generalized to $n$ photon emissions (see Supplemental Material for further details)
\begin{multline}
\label{eqn:dP_ndepsprime}
\frac{dP_n}{d\eps'}(\eps',t) =  \int_{-\infty}^{t} d\tau \, S(t,\tau;\eps') 
\\
\times \int_{\eps'}^{\eps_i} d\eps \, \frac{d^2W}{d\tau\, d\eps'}(\eps',\eps,\tau) \frac{dP_{n-1}}{d\eps}(\eps,\tau).
\end{multline}
In Eqs.~(\ref{eqn:dP_1depsprime}) and (\ref{eqn:dP_ndepsprime}) we have implicitly assumed that the local radiation probability $({d^2W}/{d\tau{}d\eps'})(\eps',\eps,\tau)$ depends only on the time $\tau$ at which the photon is emitted and on the electron instantaneous energy $\eps$. However, the position of the electron and thus the instantaneous field strength depends, in general, on the full history of previous emissions and not just on $\tau$ and $\eps$. This is a reasonable approximation when the particle is ultrarelativistic and the background field is transversely sufficiently homogeneous.

Finally, we are interested in the asymptotic probabilities $P_n$ that an electron has emitted exactly $n$ photons during the interaction
\begin{gather}
\label{eqn:Pn}
P_n = P_n(\infty),\quad P_n(t)  = \int_0^{\eps_i} d\eps' \, \frac{dP_n}{d\eps'}(\eps',t),
\end{gather}
with $P_0(t) = S(t,-\infty;\eps_i)$. In the classical limit ($\chi\ll 1$) the recoil is negligible; thus, $S(t,\tau;\eps'\approx\eps) S(\tau,t';\eps) \approx S(t,t';\eps)$, and one finds that the number of emitted photons $P_n$ follows a Poissonian distribution (see Supplemental Material and Refs.~\cite{itzykson_quantum_2005, glauber_notes_1951, dipiazzaPRL10} for further details)
\begin{gather}
\label{eqn:photonemissionprobability_glauber}
P_n = \frac{\Wglauber^n}{n!} \exp{(-\Wglauber)},
\quad
\average{n}  = \sum_{n=0}^{\infty} n P_n = \Wglauber.
\end{gather}
Here, the decay exponent $\Wglauber = \int_{-\infty}^{+\infty} d\tau\, (\nfrac{dW}{d\tau})(\eps_i,\tau)$ factorizes, is independent of the number of emitted photons, and is constant across the spectrum. This is in sharp contrast to the $\chi\gg1$ regime, where it is highly probable that $\eps' \ll \eps$ such that $S(t,\tau;\eps') S(\tau,t';\eps) \neq S(t,t';\eps)$ and the decay exponent changes substantially when multiple photon emissions become probable, which results in a qualitative change of the energy distribution even for photons originating from electrons that emitted only once [see the orange line in \figref{fig:3}(b)]. 

In \figref{fig:4} the simulated distribution (blue circles) is compared to the Poissonian prediction (orange triangles). For short interaction times (inset of \figref{fig:4}) $P_0$ is dominant, and the Poissonian approximation is valid. However, when $\chi\gg 1$ and $P_{n>0}$ is dominant, substantial deviations are found (see the main plot of \figref{fig:4}). 

Finally, we consider the attainable luminosity of a gamma-gamma collider based either on beamstrahlung in the $\chi \gg 1$ regime or on Compton backscattering. The luminosity of two identical Gaussian photon bunches colliding head-on is $\mathcal{L} = f_c N_\gamma^2/4\pi \sigma_x \sigma_y$, where $f_c$ is the bunch collision frequency, $N_\gamma$ the number of photons per bunch, and $\sigma_x$, $\sigma_y$ the rms transverse bunch sizes. Both for beamstrahlung and for Compton backscattering, high-energy photon emission occurs in a cone with $1/\gamma_p$ opening angle around the propagation direction of the emitting lepton. Hence, for the same probe electron beam and after ballistic propagation, the photon bunches generated with the two methods have similar $\sigma_x$ and $\sigma_y$ at collision. Also, $f_c$ is determined by the electron beam (and laser for Compton backscattering) repetition rate, which is assumed to be similar for both beamstrahlung and Compton backscattering. In this case, the decisive factor for achieving high luminosity is $N_\gamma$. 

In a multi-TeV gamma-gamma collider higher energy photons are of interest. For beamstrahlung in the $\chi\gg1$ regime reached with a 1~TeV probe beam and the same source beam as considered above, $\omega_{\text{min}}\approx 0.75$ with the quasimonochromatic peak at  $\omega_{\text{max}}\approx 0.998$ and more than 11\% photons with $\omega>\omega_{\text{min}}$. For Compton backscattering the highest attainable photon energy is $\omega_{\text{C,max}}=x/1+x$, where $x=4\varepsilon \varepsilon_{L}/m^2 c^4$ and $\varepsilon_{L}$ is the incident photon energy~\cite{telnovNIMA95}. In order to reach $\omega_{\text{C,max}}\approx\omega_{\text{max}}\approx 0.998$, $x\approx 500$ is required, which can be obtained by colliding $\varepsilon_{L}\approx33\text{ eV}$ photons generated by a free-electron laser with the 1~TeV probe beam. However, for $x \gtrsim 4.8$ ($x \gtrsim 8$) the competing process $\gamma \gamma_0 \rightarrow e^-e^+$ ($e^- \gamma_0 \rightarrow e^- e^- e^+$), where $\gamma$ and $\gamma_0$ refer to the Compton-backscattered and incident photon respectively, is not kinematically forbidden and has, in general, a larger cross section than the Compton one~\cite{telnovNIMA90, telnovNIMA95}. Therefore, traditional gamma-gamma collider designs focus on the regime $x \lesssim 4.8$~\cite{telnovNIMA90, telnovNIMA95}. Recently, however, research on high-energy photon production via Compton backscattering in the $x\gg1$ regime has been pursued~\cite{barklowXXX20, ginzburgXXX20}. These proposals suggest to employ polarized laser pulses colliding with polarized electron beams to suppress backgrounds. Remarkably, because of the polarization dependence of the photon emission spectrum, strongly peaked emission with photon energy around $\omega_{\text{max}}$ and drastic suppression of lower-energy photons is possible also for beamstrahlung in the $\chi\gg1$ regime by employing a polarized probe electron beam (see Fig.~4 in Ref.~\cite{seiptPRA20}).

In conclusion, we have shown that in the supercritical QED regime the beamstrahlung spectrum exhibits a quasimonochromatic peak close to the energy of the emitting lepton, and that the height of this peak provides direct quantitative \emph{in situ} information on the quantum parameter $\chi$ achieved during the interaction. Moreover, we have shown that, due to the presence of strong correlations, this peak vanishes in a regime where multiphoton emissions become dominant. The recoil-induced correlations between different photon emissions manifest themselves in a photon statistic which significantly deviates from a Poissonian distribution. These results pave the way to a photon source that is capable of efficiently delivering above TeV-energy photons, a critical step towards realizing a multi-TeV photon-photon collider.

\begin{acknowledgments}
The authors would like to thank Fabrizio Del Gaudio, Antonino Di Piazza, Gerald Dunne, Nathaniel Fisch, Thomas Grismayer, Christoph Keitel, Michael Peskin, David Reis, Lu{\'i}s Silva, Glen White, Vitaly Yakimenko, and all participants of the 2019 SLAC workshop \textit{``Physics Opportunities at a Lepton Collider in the Fully Nonperturbative QED Regime''} for stimulating discussions. At SLAC, S.M. was supported by the U.S.\ Department of Energy under contract number DE-AC02-76SF00515. At Princeton, S.M. received funding from the Deutsche Forschungsgemeinschaft (DFG, German Research Foundation) under Grant No.\ 361969338.

M.T. discovered that a peak appears in the photon emission probability in the supercritical QED regime for $\chi \gtrsim 16$ and studied its properties in 2014 and, in particular, obtained Eqs.~(\ref{minmax}) and (\ref{H}) of the main text and Eqs. (S1)-(S2c) of Supplemental Material, and performed the statistical analysis on the photon emission number distribution. S.M. independently considered the influence of the radiative decay on the single-photon emission spectrum for $\chi<10$, joined the project in 2016, and pointed out the peak suppression in the multiphoton regime. The manuscript was written equally by both authors, with simulations and figures by M.T.
\end{acknowledgments}

\bibliography{My_Bibliography}

\clearpage


\section*{Supplementary material}

# Supplemental Material for "Efficient High-Energy Photon Production in the Supercritical QED Regime"

Matteo Tamburini[1, *] and Sebastian Meuren[2, 3, †]

[1] *Max-Planck-Institut für Kernphysik, Saupfercheckweg 1, D-69117 Heidelberg, Germany*
[2] *Stanford PULSE Institute, SLAC National Accelerator Laboratory, Menlo Park, CA 94025*
[3] *Department of Astrophysical Sciences, Princeton University, Princeton, NJ 08544*
(Dated: May 22, 2021)

In the following we report a fully three dimensional (3D) analytical solution of Maxwell equations for a finite size charged beam (see Sec. I), describe the employed Monte Carlo simulation technique in more detail (see Sec. II), and discuss the details of the derivation of the multiphoton emission distribution (see Sec. III).

## I. THREE DIMENSIONAL ANALYTICAL SOLUTION OF MAXWELL EQUATIONS

In this section Gaussian units are used. State-of-the-art simulations of beam-beam interaction apply certain approximations to describe the field configuration associated with the bunch charge [1]. Here we report an analytical expression for the particle beam fields which *exactly* solves 3D Maxwell's equations. To this end we assume that in its instantaneous rest frame the electron beam is cold. For the beam, we consider the following rest frame finite size and cylindrical-symmetric charge density distribution

$$\rho(x,y,z) = \rho_0 \frac{e^{-\frac{x^2+y^2}{2\sigma^2}}}{2} \left[ \operatorname{erf}\left(\frac{\ell-2z}{2\sqrt{2}\sigma}\right) + \operatorname{erf}\left(\frac{\ell+2z}{2\sqrt{2}\sigma}\right) \right], \tag{S1}$$

where $\rho_0 = eN/(2\pi\sigma^2\ell)$ is the peak charge density, and $\operatorname{erf}(x)$ is the error function, which is given by $\operatorname{erf}(x) = (2/\sqrt{\pi})\int_0^x e^{-t^2}\,dt$. The above charge density distribution describes a finite electron beam centered at the origin, with cylindrical symmetry around the $z$ axis, transverse rms size $\sigma$ along the radial direction $r = \sqrt{x^2+y^2}$, length $\ell$ along the longitudinal axis $z$, and with $N$ particles in the beam, each with charge $e$. For the charge distribution in Eq. (S1), we found the corresponding exact analytical solution by solving Maxwell equations. The corresponding electric field components of the beam are

$$E_x(x,y,z) = \frac{eNx}{\ell(x^2+y^2)} \left[ \frac{(\ell-2z)\operatorname{erf}\left(\frac{\sqrt{(z-\frac{\ell}{2})^2+x^2+y^2}}{\sqrt{2}\sigma}\right)}{\sqrt{\ell^2-4\ell z+4(x^2+y^2+z^2)}} + \frac{(\ell+2z)\operatorname{erf}\left(\frac{\sqrt{(\frac{\ell}{2}+z)^2+x^2+y^2}}{\sqrt{2}\sigma}\right)}{\sqrt{\ell^2+4\ell z+4(x^2+y^2+z^2)}} \right.$$
$$\left. - e^{-\frac{x^2+y^2}{2\sigma^2}} \left( \operatorname{erf}\left(\frac{\ell-2z}{2\sqrt{2}\sigma}\right) + \operatorname{erf}\left(\frac{\ell+2z}{2\sqrt{2}\sigma}\right) \right) \right]; \tag{S2a}$$

$$E_y(x,y,z) = \frac{eNy}{\ell(x^2+y^2)} \left[ \frac{(\ell-2z)\operatorname{erf}\left(\frac{\sqrt{(z-\frac{\ell}{2})^2+x^2+y^2}}{\sqrt{2}\sigma}\right)}{\sqrt{\ell^2-4\ell z+4(x^2+y^2+z^2)}} + \frac{(\ell+2z)\operatorname{erf}\left(\frac{\sqrt{(\frac{\ell}{2}+z)^2+x^2+y^2}}{\sqrt{2}\sigma}\right)}{\sqrt{\ell^2+4\ell z+4(x^2+y^2+z^2)}} \right.$$
$$\left. - e^{-\frac{x^2+y^2}{2\sigma^2}} \left( \operatorname{erf}\left(\frac{\ell-2z}{2\sqrt{2}\sigma}\right) + \operatorname{erf}\left(\frac{\ell+2z}{2\sqrt{2}\sigma}\right) \right) \right]; \tag{S2b}$$

$$E_z(x,y,z) = \frac{2eN}{\ell} \left[ \frac{\operatorname{erf}\left(\frac{\sqrt{(z-\frac{\ell}{2})^2+x^2+y^2}}{\sqrt{2}\sigma}\right)}{\sqrt{\ell^2-4\ell z+4(x^2+y^2+z^2)}} - \frac{\operatorname{erf}\left(\frac{\sqrt{(\frac{\ell}{2}+z)^2+x^2+y^2}}{\sqrt{2}\sigma}\right)}{\sqrt{\ell^2+4\ell z+4(x^2+y^2+z^2)}} \right]. \tag{S2c}$$

* matteo.tamburini@mpi-hd.mpg.de

† smeuren@stanford.edu

The above field and charge density satisfy Maxwell equations exactly, i.e., $\boldsymbol{\nabla} \cdot \boldsymbol{E} = 4\pi\rho$ and $\boldsymbol{\nabla} \times \boldsymbol{E} = 0$, while $\boldsymbol{B} = 0$ in the rest frame of the beam.

The fields of a beam moving along the positive $z$ axis with velocity $v$ and relativistic factor $\gamma$ are found via a Lorentz transformation of the rest frame fields

$$E_x^{\text{lab}}(x,y,z,t) = \gamma E_x^{\text{rest}}(x,y,\gamma(z-vt)), \quad \text{(S3a)}$$
$$E_y^{\text{lab}}(x,y,z,t) = \gamma E_y^{\text{rest}}(x,y,\gamma(z-vt)), \quad \text{(S3b)}$$
$$E_z^{\text{lab}}(x,y,z,t) = E_z^{\text{rest}}(x,y,\gamma(z-vt)), \quad \text{(S3c)}$$
$$B_x^{\text{lab}}(x,y,z,t) = -\gamma v E_y^{\text{rest}}(x,y,\gamma(z-vt))/c, \quad \text{(S3d)}$$
$$B_y^{\text{lab}}(x,y,z,t) = \gamma v E_x^{\text{rest}}(x,y,\gamma(z-vt))/c, \quad \text{(S3e)}$$
$$B_z^{\text{lab}}(x,y,z,t) = 0, \quad \text{(S3f)}$$

where the superscript lab (rest) denote the laboratory (rest frame) fields. One can explicitly verify that the above fields satisfy the full set of Maxwell equations exactly

$$\boldsymbol{\nabla} \cdot \boldsymbol{E}^{\text{lab}} = 4\pi\rho^{\text{lab}}, \quad \text{(S4a)}$$
$$\boldsymbol{\nabla} \cdot \boldsymbol{B}^{\text{lab}} = 0, \quad \text{(S4b)}$$
$$\boldsymbol{\nabla} \times \boldsymbol{E}^{\text{lab}} + \frac{\partial \boldsymbol{B}^{\text{lab}}}{\partial ct} = 0, \quad \text{(S4c)}$$
$$\boldsymbol{\nabla} \times \boldsymbol{B}^{\text{lab}} - \frac{\partial \boldsymbol{E}^{\text{lab}}}{\partial ct} = \frac{4\pi}{c} \rho^{\text{lab}} v, \quad \text{(S4d)}$$
$$\rho^{\text{lab}}(x,y,z,t) = \gamma \rho^{\text{rest}}(x,y,\gamma(z-vt)). \quad \text{(S4e)}$$

## II. DESCRIPTION OF THE MONTE-CARLO APPROACH

In this section Gaussian units are used. Our implementation of stochastic photon emission by an electron or a positron as well as the creation of an electron-positron pair from an energetic photon in the presence of a strong-background field are detailed in Ref. [2]. Here we summarize the main steps. At each time step $\Delta t$ the quantum parameter $\chi$ is calculated according to

$$\chi = \frac{|e|\hbar}{m^3 c^4} \sqrt{(\varepsilon \mathbf{E}/c + \mathbf{p} \times \mathbf{B})^2 - (\mathbf{p} \cdot \mathbf{E})^2}, \quad \text{(S5)}$$

where $\varepsilon$ and $\mathbf{p}$ are the particle energy and momentum, respectively. The total probability of emitting a photon

$$\frac{dW_{\text{rad}}}{dt} = \int_0^{\varepsilon} \frac{d^2 W_{\text{rad}}(\varepsilon_\gamma)}{dt d\varepsilon_\gamma} d\varepsilon_\gamma \quad \text{(S6)}$$

is calculated, where $d^2 W_{\text{rad}}(\varepsilon_\gamma)/dt d\varepsilon_\gamma$ is the differential probability for an electron/positron with energy $\varepsilon$ to emit a photon with energy $\varepsilon_\gamma$ (see Eq. (4.24) in Ref. [3])

$$\frac{d^2 W_{\text{rad}}}{dt d\varepsilon_\gamma} = \frac{\alpha m^2 c^4}{\sqrt{3}\pi \hbar \varepsilon^2} \frac{1}{(1+u)} \left\{ [1 + (1+u)^2] K_{2/3}\left(\frac{2u}{3\chi}\right) \right.$$
$$\left. - (1+u) \int_{2u/3\chi}^{\infty} K_{1/3}(y) dy \right\}, \quad \text{(S7)}$$

where $u = \varepsilon_\gamma/(\varepsilon - \varepsilon_\gamma)$ and $K_\nu(x)$ are the modified Bessel functions of second kind. For each electron and positron a photon emission occurs if $r_1 < \Delta t dW_{\text{rad}}/dt$, where $0 < r_1 < 1$ is a uniformly distributed random number being generated at each time step. The time step is chosen such that the condition $\Delta t dW_{\text{rad}}/dt \ll 1$ holds.

If the above-mentioned condition is fulfilled, the energy of the emitted photon $\varepsilon_\gamma$ is obtained as the root of the sampling equation

$$\int_0^{\varepsilon_\gamma} \frac{d^2 W_{\text{rad}}}{dt d\varepsilon_\gamma} d\varepsilon_\gamma = r_2 \frac{dW_{\text{rad}}}{dt}, \quad \text{(S8)}$$

where $0 < r_2 < 1$ is a uniformly distributed random number, independent of $r_1$, that is generated at each photon-creation event. The direction of propagation of the emitted photon is parallel to the momentum $\mathbf{p}$ of the parental particle, as emission occurs within a cone with an opening angle of the order of $1/\gamma$ [3]. The event generator for the creation of an electron-positron pair from a photon follows the same steps as for photon emission, except that $d^2 W_{\text{rad}}/dt d\varepsilon_\gamma$ is replaced by the differential probability for a photon with energy $\varepsilon_\gamma$ to convert into an electron-positron pair $d^2 W_{\text{pair}}/dt d\varepsilon$, which is given by (see Eq. (3.50) in Ref. [3])

$$\frac{d^2 W_{\text{pair}}}{dt d\varepsilon_e} = \frac{\alpha m^2 c^4}{\sqrt{3}\pi \hbar \varepsilon_\gamma^2} \left[ \frac{\varepsilon_e^2 + \varepsilon_p^2}{\varepsilon_e \varepsilon_p} K_{2/3}\left(\frac{2\varepsilon_\gamma^2}{3\chi_\gamma \varepsilon_e \varepsilon_p}\right) + \int_{2\varepsilon_\gamma^2/3\chi_\gamma \varepsilon_e \varepsilon_p}^{\infty} dy K_{1/3}(y) \right], \quad \text{(S9)}$$

where $\chi_\gamma$ is the quantum parameter of the photon, $\varepsilon_e$ is the electron energy ($0 < \varepsilon_e < \varepsilon_\gamma$) and $\varepsilon_p = (\varepsilon_\gamma - \varepsilon_e)$ is the positron energy.

## III. DERIVATION OF THE MULTIPHOTON EMISSION DISTRIBUTION

Assuming that the locally-constant-field approximation holds, the single-photon emission probability [3]

$$\frac{d^2 W}{dt d\omega} = \frac{\alpha}{\sqrt{3}\pi \tau_c \gamma} \left\{ \left[ 2 + \frac{\omega^2}{(1-\omega)} \right] K_{2/3}\left[\frac{2\omega}{3\chi(1-\omega)}\right] \right.$$
$$\left. - \int_{2\omega/[3\chi(1-\omega)]}^{\infty} dy\, K_{1/3}(y) \right\}, \quad \text{(S10)}$$

is always applicable for sufficiently small time intervals $dt$. Here $\tau_c = \hbar/(mc^2)$ is the Compton time and $K_\nu(x)$ is the modified Bessel function of second kind [4].

As the electron energy after emission $\varepsilon'$ is determined by the initial electron energy $\varepsilon$ and the normalized emitted photon energy $\omega = \varepsilon_\gamma/\varepsilon$ via $\varepsilon' = \varepsilon - \varepsilon_\gamma = \varepsilon(1-\omega)$,

we obtain the relation

$$\frac{dW}{dt}(\varepsilon,t) = \int_0^\varepsilon d\varepsilon' \frac{d^2W}{dtd\varepsilon'}(\varepsilon',\varepsilon,t)$$
$$= -\int_0^\varepsilon d\varepsilon' \frac{1}{\varepsilon}\frac{d^2W}{dtd\omega}[\omega(\varepsilon',\varepsilon),\varepsilon,t]. \quad (S11)$$

In the following we focus on the final electron energy spectrum (the derivation of the emitted photon spectrum is completely analogous). Time integrals are calculated along the electron trajectory, and we implicitly assume that the work performed by the external field is negligible. This is rigorously exact, e.g., for magnetic fields.

We start with the "survival probability" $S(t,t';\varepsilon)$, i.e., the probability that an electron with energy $\varepsilon$ does *not* emit a photon during the time interval $[t,t']$. Firstly, we note that this probability should obey

$$S(t,\tau;\varepsilon)S(\tau,t';\varepsilon) = S(t,t';\varepsilon), \quad S(t,t;\varepsilon) = 1. \quad (S12)$$

Furthermore, it represents a solution to the differential equation

$$\frac{d}{dt}S(t,t';\varepsilon) = -\frac{dW}{dt}(\varepsilon,t)S(t,t';\varepsilon). \quad (S13)$$

Thus, we obtain an exponential decay factor

$$S(t,t';\varepsilon) = \exp\left[-\int_{t'}^t d\tau \frac{dW}{d\tau}(\varepsilon,\tau)\right], \quad (S14)$$

and the total emission probability per unit time $dW(\varepsilon,t)/dt$ [see Eq. (S11)] defines the electron radiative lifetime (see [5–7] for a QFT derivation based on the radiatively corrected electron wavefunction).

It is useful to define the quantity $(dP_n/d\varepsilon')(\varepsilon',t)$, which denotes the probability that an electron with initial energy $\varepsilon_i$ (at $t \to -\infty$) has emitted exactly $n$ photons during the time interval $[-\infty,t]$ and its final energy is within the infinitesimal range $[\varepsilon',\varepsilon'+d\varepsilon']$. This quantity is recursively given by

$$\frac{dP_n}{d\varepsilon'}(\varepsilon',t) = \int_{-\infty}^t d\tau\, S(t,\tau;\varepsilon')$$
$$\times \int_{\varepsilon'}^{\varepsilon_i} d\varepsilon \frac{d^2W}{d\tau d\varepsilon'}(\varepsilon',\varepsilon,\tau)\frac{dP_{n-1}}{d\varepsilon}(\varepsilon,\tau). \quad (S15)$$

As the $n$th emission may happen at any time $\tau$ within $[-\infty,t]$, one has to integrate over the emission time $\tau$ and multiply the probabilities that there were a) exactly $n-1$ emissions until $\tau$, b) an emission at $\tau$ and c) zero emissions between $\tau$ and $t$. Similarly, the emitted energy can be arbitrarily split between the last and the $n-1$ emissions before. Thus, one also has to integrate over $\varepsilon$ in Eq. (S15), which denotes the electron energy after the first $n-1$ emissions.

As the electron has a definite initial energy $\varepsilon_i$, we find

$$\frac{dP_1}{d\varepsilon'}(\varepsilon',t) = \int_{-\infty}^t d\tau\, S(t,\tau;\varepsilon')$$
$$\times \frac{d^2W}{d\tau d\varepsilon'}(\varepsilon',\varepsilon_i,\tau)S(\tau,-\infty;\varepsilon_i), \quad (S16)$$

which serves as starting point for the recursion specified in Eq. (S15).

### A. Classical limit

Next, we focus on the asymptotic probabilities $P_n$ that an electron emits in total $n$ photons during the interaction

$$P_n = P_n(\infty), \quad P_n(t) = \int_0^{\varepsilon_i} d\varepsilon' \frac{dP_n}{d\varepsilon'}(\varepsilon',t) \quad (S17)$$

$[P_0(t) = S(t,-\infty;\varepsilon_i)]$ and show that in the classical limit, i.e., $\chi \ll 1$ such that $\varepsilon' \approx \varepsilon$, these probabilities follow a Poissonian distribution [8, 9]

$$P_n = \frac{\mathcal{W}^n}{n!}\exp(-\mathcal{W}), \quad \langle n \rangle = \sum_{n=0}^\infty nP_n = \mathcal{W}, \quad (S18)$$

where $\exp(-\mathcal{W}) = S(\infty,-\infty;\varepsilon_i)$ with [see Eq. (S14)]

$$\mathcal{W} = \int_{-\infty}^{+\infty} d\tau \frac{dW}{d\tau}(\varepsilon_i,\tau). \quad (S19)$$

Note that in an $S$-matrix–based derivation of Eq. (S18) the appearance of the nonperturbative decay exponent $\exp(-\mathcal{W})$ is a consequence of radiative corrections [8, 9]. In a QFT calculation $\exp(-\mathcal{W}) = S(\infty,-\infty;\varepsilon_i)$ appears as soon as radiative corrections to the electron/positron wave function are taken into account. Therefore, radiative corrections to tree-level diagrams need to be included when $\exp(-\mathcal{W})$ significantly deviates from unity (see [5–7, 10]).

As the recoil is negligible in the classical regime, only electron energies $\varepsilon' \approx \varepsilon$ contribute significantly to $(d^2W/d\tau\,d\varepsilon')(\varepsilon',\varepsilon,\tau)$. Hence, we can replace all decay exponents $S(t,t';\varepsilon')$ in Eq. (S15) and Eq. (S16) with $S(t,t';\varepsilon_i)$. Thus, we inductively find

$$\frac{dP_n}{d\varepsilon'}(\varepsilon',t) = S(t,-\infty;\varepsilon_i)\frac{d\widetilde{P}_n}{d\varepsilon'}(\varepsilon',t), \quad (S20)$$

where

$$\frac{d\widetilde{P}_n}{d\varepsilon'}(\varepsilon',t) = \int_{-\infty}^t d\tau \frac{d^2W}{d\tau\,d\varepsilon'}(\varepsilon',\varepsilon_i,\tau)\,\widetilde{P}_{n-1}(\tau), \quad (S21a)$$

$$\frac{d\widetilde{P}_1}{d\varepsilon'}(\varepsilon',t) = \int_{-\infty}^t d\tau \frac{d^2W}{d\tau d\varepsilon'}(\varepsilon',\varepsilon_i,\tau), \quad (S21b)$$

and

$$\widetilde{P}_n(t) = \int_0^{\varepsilon_i} d\varepsilon \frac{d\widetilde{P}_n}{d\varepsilon}(\varepsilon,t). \quad (S21c)$$

Furthermore, $P_n = \exp(-\mathcal{W})\widetilde{P}_n(\infty)$ with $\exp(-\mathcal{W}) = S(\infty,-\infty;\varepsilon_i)$ and we obtain

$$\frac{d}{dt}\widetilde{P}_n(t) = \frac{dW}{dt}(\varepsilon_i,t)\widetilde{P}_{n-1}(t). \quad (S22)$$

As this system of differential equations is solved by

$$\widetilde{P}_n(t) = \frac{1}{n!} \left[ \int_{-\infty}^{t} d\tau \, \frac{dW}{d\tau}(\varepsilon_i, \tau) \right]^n, \quad (S23)$$

we recover Eq. (S18) in the limit $t \to \infty$.

### B. Probability conservation

Finally, we verify that probability is conserved at all times, i.e., that

$$\sum_{n=0}^{\infty} P_n(t) = 1. \quad (S24)$$

This follows from the time derivative

$$\frac{d}{dt} P_n(t) = \int_0^{\varepsilon_i} d\varepsilon \, \frac{dW}{dt}(\varepsilon, t)$$
$$\times \left[ \frac{dP_{n-1}}{d\varepsilon}(\varepsilon, t) - \frac{dP_n}{d\varepsilon}(\varepsilon, t) \right], \quad (S25)$$

which is obtained from Eq. (S15) after applying the integral identity

$$\int_0^{\varepsilon_i} d\varepsilon' \int_{\varepsilon'}^{\varepsilon_i} d\varepsilon \, f(\varepsilon', \varepsilon) = \int_0^{\varepsilon_i} d\varepsilon \int_0^{\varepsilon} d\varepsilon' \, f(\varepsilon', \varepsilon). \quad (S26)$$

Eq. (S25) implies that all terms cancel pairwise. Note that the derivatives of $P_0(t) = S(t, -\infty; \varepsilon_i)$ and $P_1(t)$ [see Eq. (S16)] have to be calculated explicitly.



\end{document}